\documentclass[aps,prl,amsmath,amssymb,reprint,superscriptaddress,showkeys]{revtex4-2}

\usepackage{graphicx}
\usepackage{xcolor}
\usepackage{amsmath}
\usepackage{amssymb}
\usepackage{braket}
\usepackage{subfigure}
 
\newcommand{\blue}[1]{\textcolor{blue}{#1}}
 
\begin{document}

\title{Controlling Quantum Chaos: Optimal Coherent Targeting}

\newcommand{\RegensburgUniversity}{Institut f\"ur Theoretische Physik, Universit\"at Regensburg, D-93040 Regensburg, Germany}
\newcommand{\Steve}{Department of Physics and Astronomy, Washington State University, Pullman, WA USA}

\author{Steven Tomsovic}
\email{tomsovic@wsu.edu}
\affiliation{\RegensburgUniversity}
\affiliation{\Steve}
\author{Juan Diego Urbina}
\affiliation{\RegensburgUniversity}
\author{Klaus Richter}
\affiliation{\RegensburgUniversity}

\begin{abstract}

One of the principal goals of controlling classical chaotic dynamical systems is known as targeting, which is the very weakly perturbative process of using the system's extreme sensitivity to initial conditions in order to arrive at a predetermined target state.  It is shown that a generalization to chaotic quantum systems is possible in the semiclassical regime, but requires tailored perturbations whose effects must undo the dynamical spreading of the evolving quantum state.  The procedure described here is applied to initially minimum uncertainty wave packets in the quantum kicked rotor, 
a preeminent quantum chaotic paradigm, to illustrate the method, and investigate its accuracy.  The method's error can be made to vanish as $\hbar \rightarrow 0$.

\end{abstract}



\maketitle

One of the several meanings ascribed to the term {\bf \it controlling chaos}~\cite{Ott90, Ott06} is the concept of targeting~\cite{Shinbrot90, Kostelich93, Bollt95, Schroer97} in which the exponential instability of a chaotic dynamical system is taken advantage of in an optimal way.  In Refs.~\cite{Shinbrot90, Kostelich93} small perturbations of a system parameter leading to some predetermined final state were discussed, but a very weak perturbation applied to the system's trajectory (initial conditions) can be applied as well.  A well known early example is the redirecting in late 1983 of the International Sun-Earth Explorer-3 spacecraft towards a rendezvous with the Giacobini-Zimmer comet in 1985~\cite{Farquhar85,Ott06}.  

A natural question arises as to whether it is possible to bring this concept from the classical to the quantum realm and to achieve some form of targeting with highly excited or far-from-equilibrium quantum chaotic systems.  In this regard, three desiderata can serve as a blueprint.  First, let us assume that the quantum system is isolated and has a well defined classical analog.  Ideally, the initial quantum and target states of interest would be as classical as possible given uncertainty relations.  Second, an optimal quantum transport path can be identified, conceivably guided by the classical analog's (chaotic) dynamics.  Finally, it is preferable to avoid monitoring or measurements, and to seek applying purely coherent unitary weak perturbations to the quantum system's dynamics in order to achieve the optimal targeting.  Clearly, such an \textit{\textbf{optimal coherent chaotic quantum targeting}} procedure would fall into the general classification of a quantum control problem.

The subject of quantum control theory has a long history; a survey of the theory and applications can be found in~\cite{Dong10} and a more recent pedagogical overview of optimal control theory in~\cite{James21}.  Within this enormous body of work, are some concepts a bit closer to the subject of this Letter.  In~\cite{Sugawara03}, Sugawara describes `wave packet shaping', which is applied to the integrable Morse oscillator potential.  It gives a complicated time-dependent laser field that shifts a wave packet from its ground state location to another desired location. Another method makes use of phase space structural implications of Kolmogorov-Arnol'd-Moser theory~\cite{Kolmogorov54, Arnold63, Moser62}; see Ref.~\cite{Madronero06}, which describes the creation of a non-dispersive electronic wave packet following a trajectory about which the local phase space dynamics has an approximately harmonic nature.  However, the highly excited, strongly chaotic regime remains challenging.

\begin{figure}[ht]
\includegraphics[width=\columnwidth]{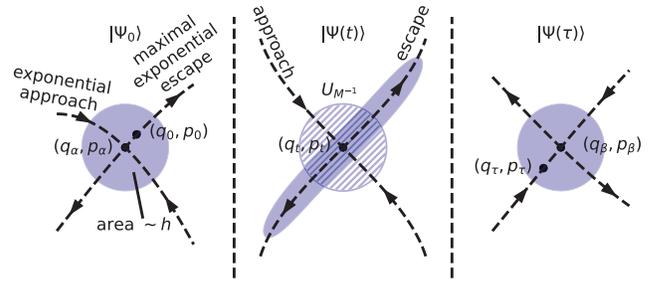}
\caption{{\bf Schematic of optimal coherent chaotic quantum targeting}.  The circular zone represents the Wigner transform density of a nearly minimal uncertainty state, which can be slightly shifted (via a unitary operator, $U_s$) to be centered on an optimal trajectory starting at $(q_0,p_0)$.  As it propagates the density follows an optimal chaotic trajectory, but is locally spreading, which must be counteracted by  contractions, $U_{M^{-1}}$. At the end, it can be shifted from $(q_\tau,p_\tau)$ to the centroid of the target state $(q_\beta,p_\beta)$.
\label{fig1}}
\end{figure}

Returning to the first desideratum, for a quantum chaotic system the initial and target states ideally would be minimum uncertainty states with well defined mean momenta and positions, $(q_\alpha, p_\alpha)$ and $(q_\beta, p_\beta)$, respectively.  The system would be in a semiclassical regime, i.e.~$\hbar$ is sufficiently small that an asymptotic theory, such as advanced methods applicable to chaotic systems~\cite{Huber88, Pal16} and based on time-dependent Wentzel-Kramers-Brillouin (WKB) theory~\cite{Maslov81, Keller58, Wentzel26, Kramers26, Brillouin26} describe the propagation well.  The Wigner transforms~\cite{Wigner32} of these states are analogous to localized Liouvillian phase space densities with volumes determined by $\hbar$ as opposed to single trajectories and this adds an inescapable new ingredient, arguably the most important one.  

As sketched in Fig.~\ref{fig1}, these densities provide a geometric picture for the task of identifying an optimal quantum transport path, the second desideratum.  For a fully chaotic Hamiltonian dynamical system, there exist directions of maximal exponential deviation under perturbations as well as maximal exponential compression~\cite{Poincare99}.  Intuitively, separating along the direction of maximal exponential rate from the centroid of the initial density while simultaneously converging at the maximal exponential rate towards the centroid of the target density, gives an optimal dynamical path from the initial to target state.  For the desired arrival time, $\tau$, there exists such a {\it heteroclinic} trajectory segment that
maximizes it's initial conditions', $(q_0, p_0)$, weight in the Liouvillian density of the initial state, and similarly for it's endpoint, $(q_\tau, p_\tau)$, within the density for the target state.  Crudely speaking, $(q_0, p_0)$ and $(q_\tau, p_\tau)$ are closest to $(q_\alpha, p_\alpha)$ and $(q_\beta, p_\beta)$, respectively.

The methods for finding heteroclinic trajectory segments developed in~\cite{Li17, Tomsovic93, Tomsovic18b} are sufficient for the purposes here, and it has previously been shown that for chaotic systems the full set of heteroclinic trajectory segments can be used to construct the full time evolution of transport coefficients involving wave packets~\cite{Tomsovic91b, Oconnor92, Tomsovic93} and coherent states in chaotic many-body bosonic systems~\cite{Tomsovic18, Tomsovic18b}.  However, to provide an optimal transport pathway for the quantum dynamics, here the idea is to follow just one segment, thus evoking the third desideratum.  Centering the density on the trajectory segment and maintaining its local nature is sufficient to accomplish this.  Both operations, translations and dilations, fall into the class of linear canonical transformations, which have precise unitary transformation counterparts as was known very early on to Dirac~\cite{Dirac30}, and was developed fully later~\cite{Collins70, Moshinsky71}.  There it was shown that these unitary transformations have at most quadratic generators.

A unitary shift transformation, say $U_s$, slightly shifting the initial and final state's centroids towards the initial and final conditions of the heteroclinic trajectory segment, is given by $U_s(\Delta p,\Delta q)= \exp\left[i\left(\Delta p \hat q - \hat p\Delta q\right)/\hbar\right]$.  In configuration space, it is expressed as
\begin{equation}
\label{shift}
\braket{q |U_s(\Delta p,\Delta q)|q'} =\! \exp\!\left[\frac{i}{\hbar}\Delta p\! \left(q\!-\!\frac{\Delta q}{2}\right) \right]\!\delta\!\left(q\!-\!q'\!-\!\Delta q\right)
\end{equation}
where the global phase part, $-\Delta p\Delta q/(2\hbar)$, is most likely of no concern and can be dropped.  Curiously, small perturbations to guide the system toward following a heteroclinic trajectory segment is a critical part of classical targeting, but is not nearly as essential in the quantum context, especially in the very strongly chaotic limit where no transport barriers are present.

Countering the spreading of the quantum state under evolution is the most critical operation and must be done periodically before the state is spread to such an extent that the nonlinear dynamics becomes manifest in the shape of the spreading Liouvillian density.   The heteroclinic trajectory segment's stability matrix for each such time interval describes the successive wave packet spreadings, and is describing a local linear canonical transformation.  The local unitary transformation associated with the inverse stability matrix for that portion of the trajectory, say $U_{M^{-1}}$, unwinds the spreading and returns the wave packet to its minimum uncertainty form. 

Within the semiclassical propagation method called linearized wave packet dynamics~\cite{Heller75, Heller91} are an accounting for this spreading, and the method effectively constructs the corresponding unitary transformation in a configuration representation.  The multidimensional expressions can be found in~\cite{Pal16}, and for an alternative approach for suppression of spreading through periodic nonlinear kicking see~\cite{Goussev18}.  The main ingredients are the second derivatives of the generating function, $S$, which can be expressed in terms of the stability matrix elements (block ordering is consistent with the kicked rotor equations ahead)~\cite{Heller91}.  To unwind this spreading using the stability matrix follows using block $2$ x $2$ matrix inversion formulas~\cite{Lu02} and some algebra.  One finds
\begin{align}
&\left(\!\begin{array}{cc}  \frac{\partial^2 S}{\partial \vec{q} \partial \vec{q}} & \frac{\partial^2 S }{\partial \vec{q}' \partial \vec{q}} \\ \frac{\partial^2 S}{\partial \vec{q} \partial \vec{q}'} & \frac{\partial^2 S }{\partial \vec{q}' \partial \vec{q}'} \end{array}\!\right) = \nonumber \\
&\left(\!\begin{array}{cc} -\bf{M^{-1}_{21}}\cdot  \bf{M_{22}} &  \bf{M^{-1}_{21}} \\ 
 \bf{M_{11}} \cdot  \bf{M^{-1}_{21}} \cdot  \bf{M_{22}} \!-\!\bf{M_{12}} \quad & \!\! -\bf{M_{11}} \cdot \bf{M^{-1}_{21}}  \end{array} \!\right) \, .
\end{align}
At the moment in time, $t$, that the spreading is countered, the transformation must be centered locally about the heteroclinic trajectory's location $(q_t,p_t)$ using the coordinates $q-q_t$ and $q'-q_t$.  In addition, centering the momentum coordinate in configuration space is accomplished by multiplying the transformation by $\exp\left[i p_t\left(q-q'\right)/\hbar\right]$.

If the unitary dynamics of the uncontrolled quantum chaotic system of interest is denoted by $U(t)$, then by creating a controlled quantum dynamics given by 
\begin{equation}
\label{unt}
U_{\rm cqd}(\tau) = U_s \prod_{j=1}^n\{U_{M^{-1}}U(t)\}_jU_s\, ,
\end{equation}
with $U_{M^{-1}}$ unwinding spreading, $\tau=nt$, and the shift operators $U_s$,
Eq.~(\ref{shift}), the initial quantum state can be made to follow any trajectory segment that exists in the classical analog dynamical system.  This technique of optimal coherent chaotic quantum targeting is illustrated for the quantum kicked rotor next. 

\begin{figure}[ht]
 \begin{subfigure}
  \centering
  \includegraphics[width=\columnwidth]{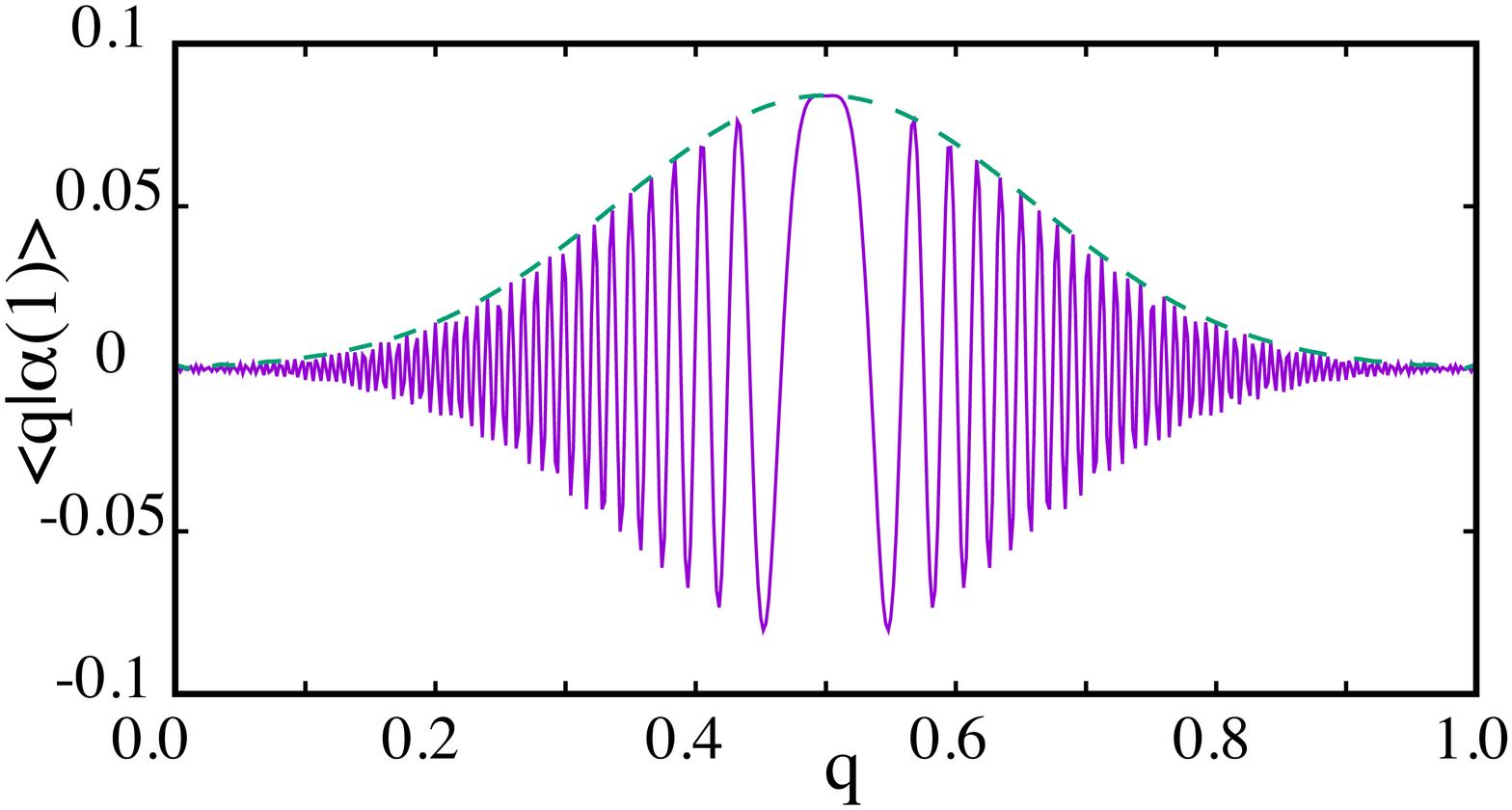}
  \label{fig2a}
 \end{subfigure}
 \begin{subfigure} 
  \centering
  \includegraphics[width=\columnwidth]{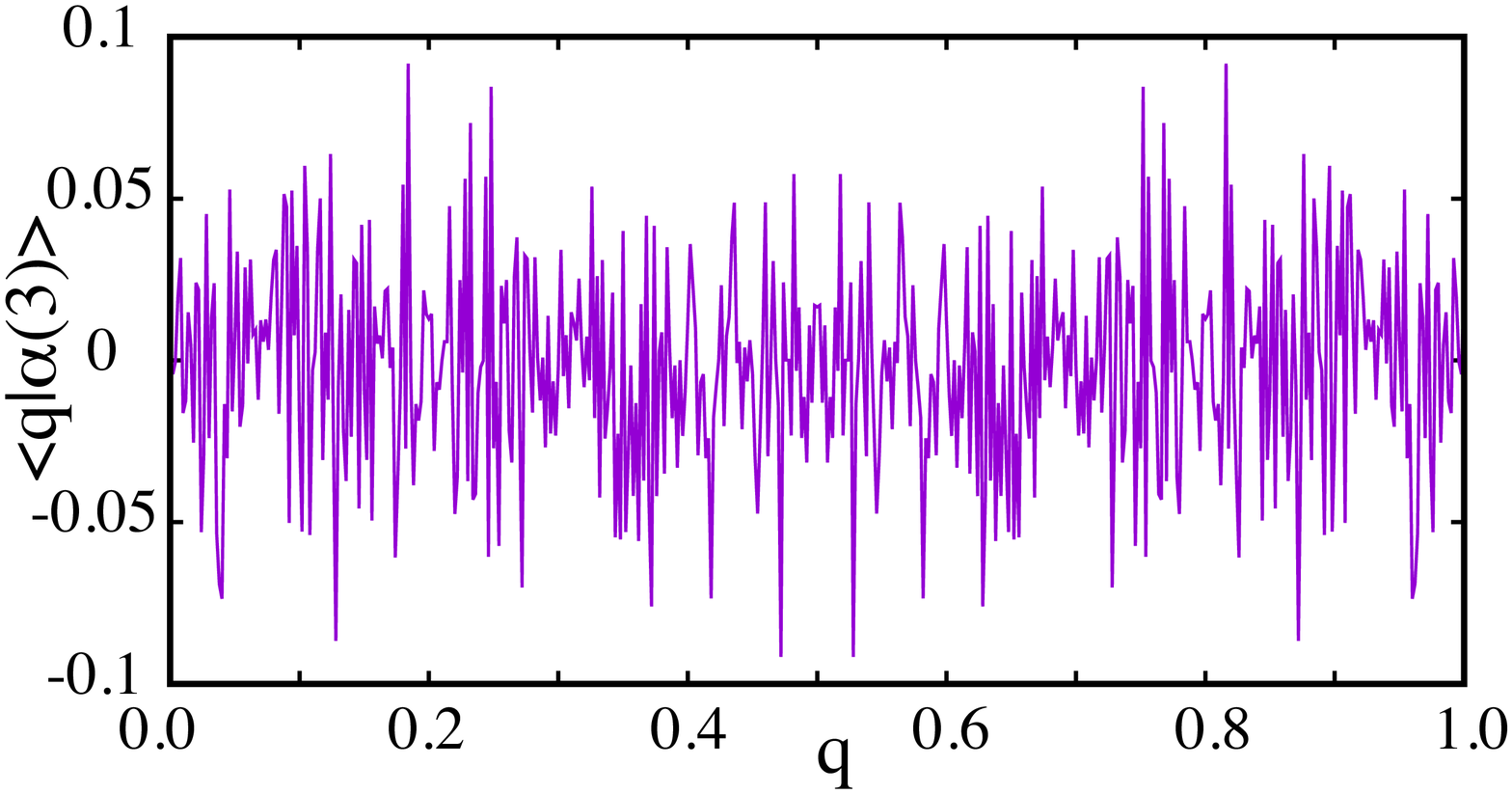}
  \label{fig2b}
 \end{subfigure}
\caption{{\bf Quantum chaotic spreading of the kicked rotor initial state $|\alpha\rangle$}.  The real (purple solid) and absolute (dashed green) values of  the one and three time propagations of the initial state, $\braket{q|\alpha(1) }$ (top panel) and $\braket{q|\alpha (3) }$ (bottom panel).  If uncontrolled the state spreads and rapidly acquires an ergodic looking appearance.}
\label{fig2}
\end{figure}

\begin{figure}[ht]
    \begin{subfigure}
            \centering
            \includegraphics[width=8.1cm]{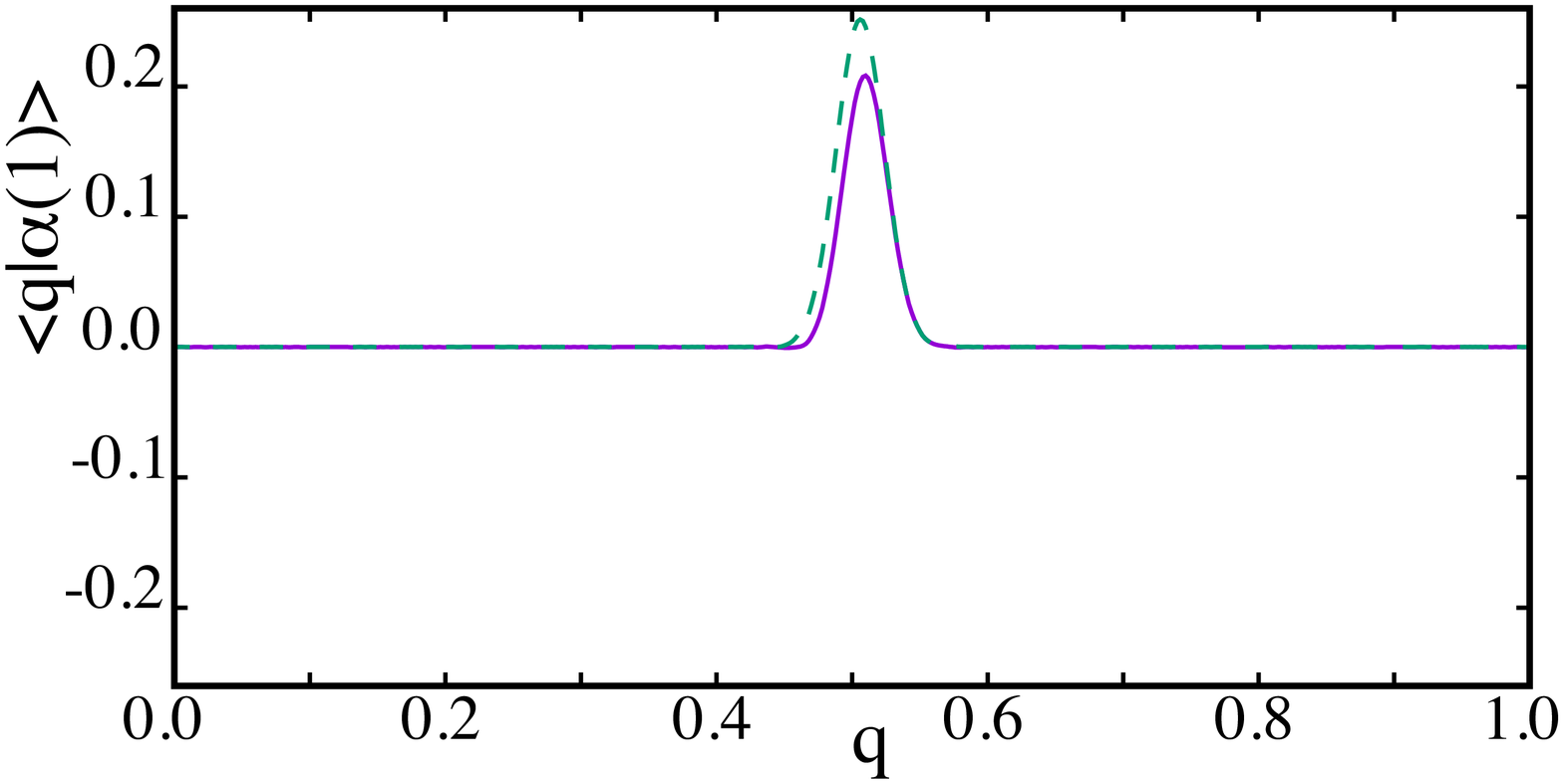}
            \label{fig3a}
    \end{subfigure}
    \begin{subfigure}
            \centering
            \includegraphics[width=8.1cm]{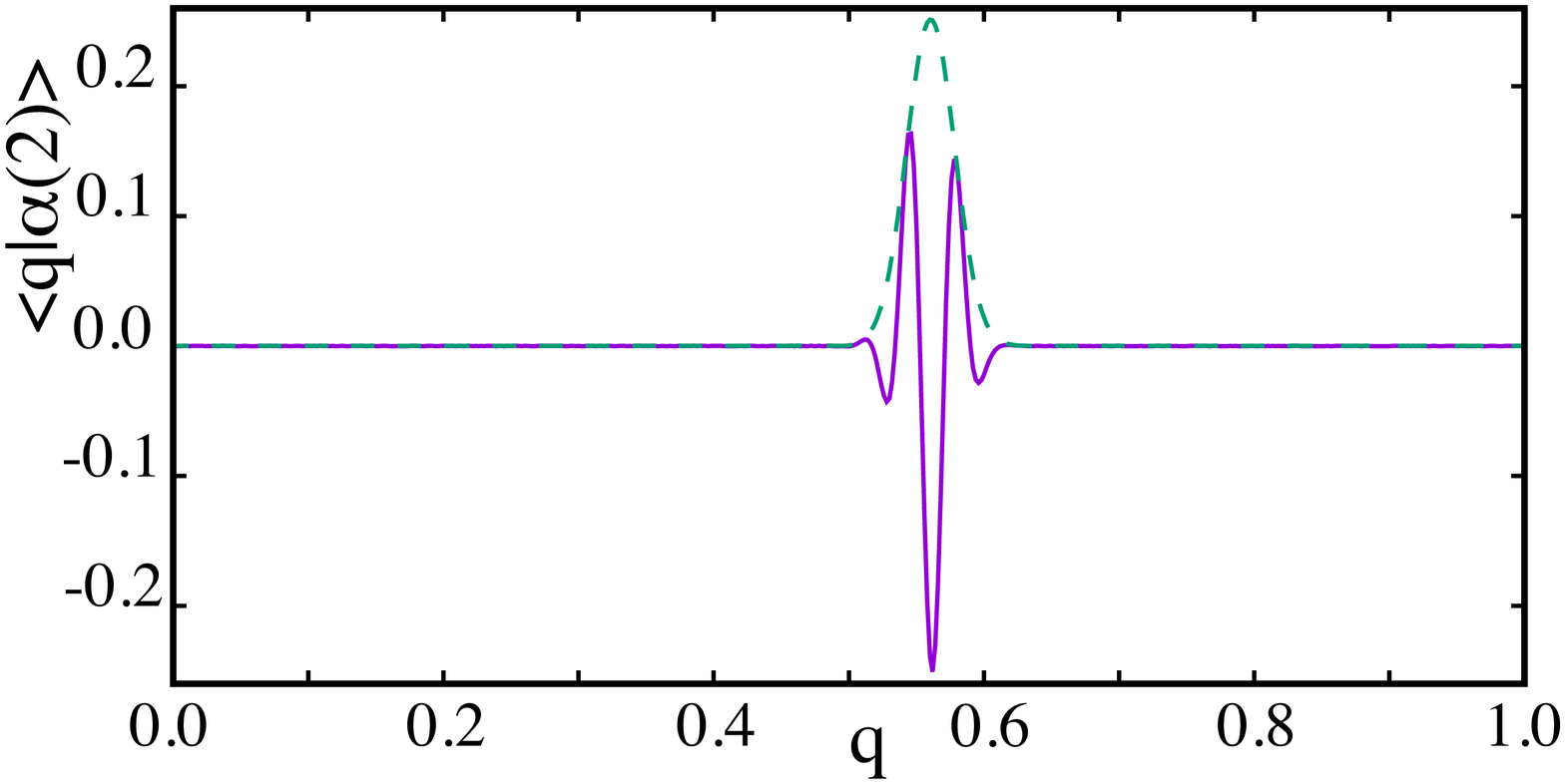}
            \label{fig3b}
    \end{subfigure}
    \begin{subfigure} 
            \centering
            \includegraphics[width=8.1cm]{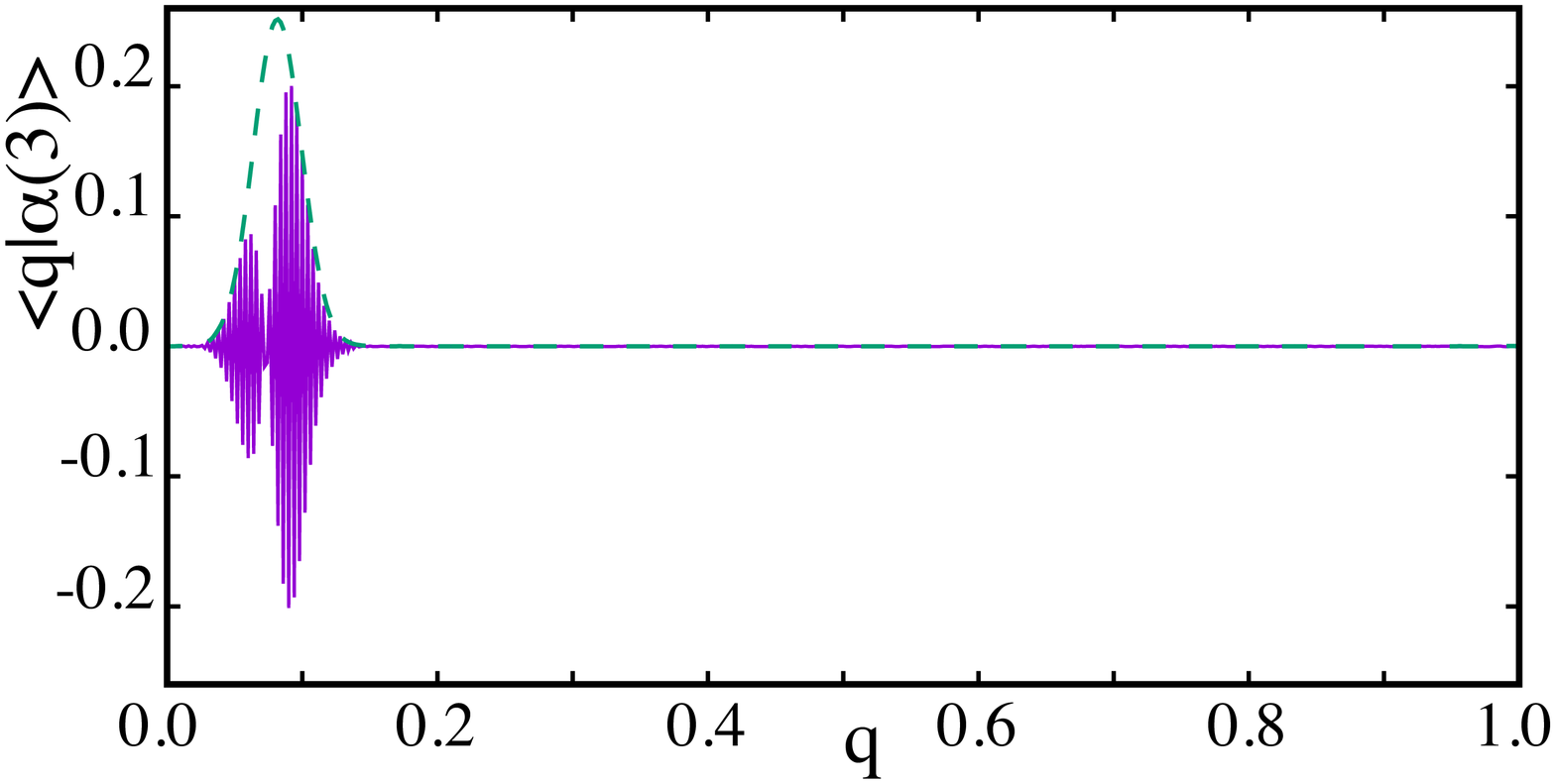}
            \label{fig3c}
    \end{subfigure}
    \begin{subfigure} 
            \centering
            \includegraphics[width=8.1cm]{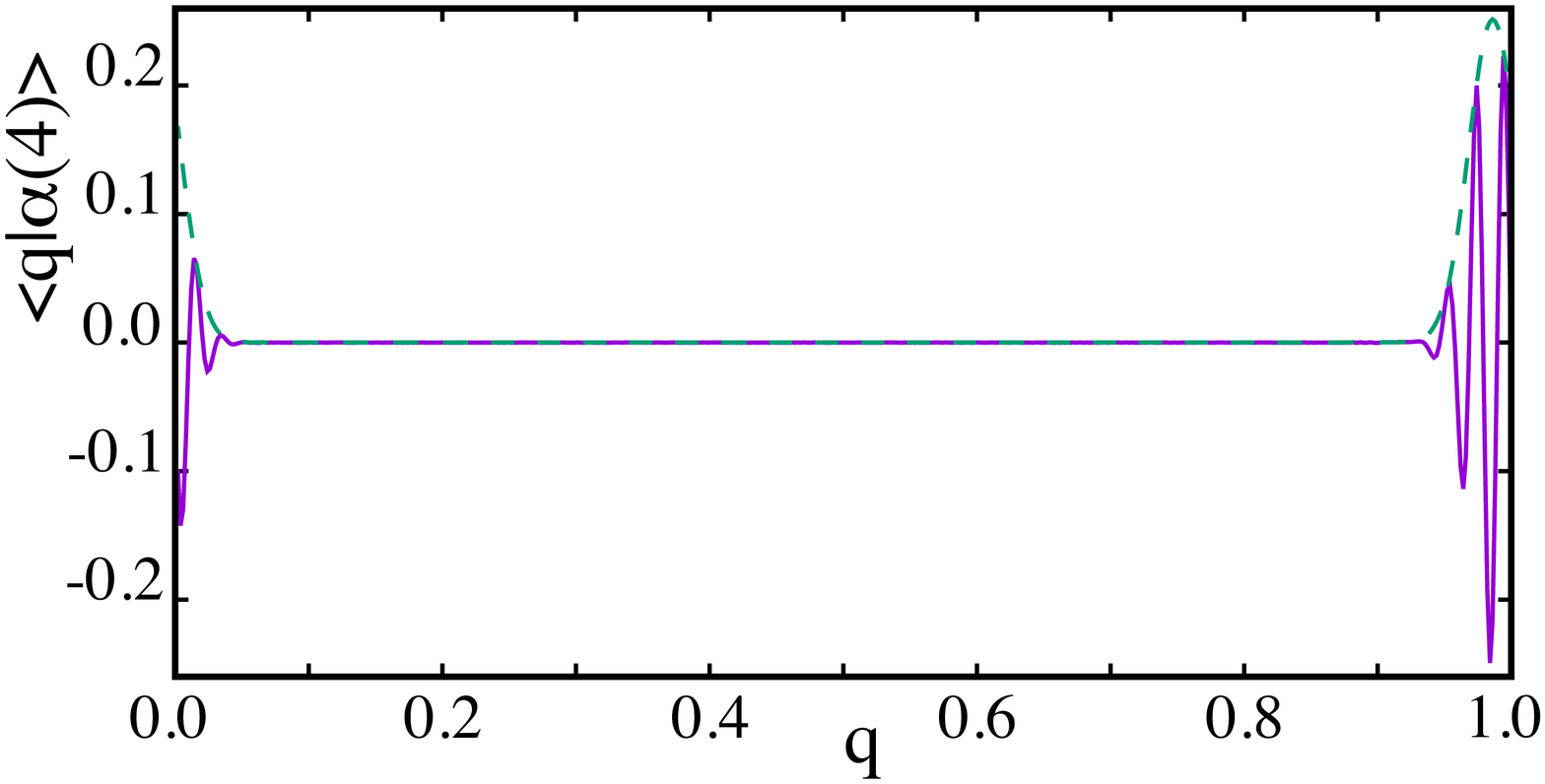}
            \label{fig3d}
    \end{subfigure}
    \caption{{\bf Quantum guiding along heteroclinic classical paths}.  The real (solid purple) and absolute (dashed green) values of $\braket{q|\alpha(t)}$ are obtained using Eq.~\eqref{unt}.  At each step the propagated state has a near perfect overlap (not discernible directly from the figure) with a wave packet of the form of Eq.~\eqref{gauss} with the time appropriate point of the heteroclinic trajectory in Table~\ref{table1}.
     }
\label{fig3}
\end{figure}

\begin{figure}[ht]
\centering
\includegraphics[width=\columnwidth]{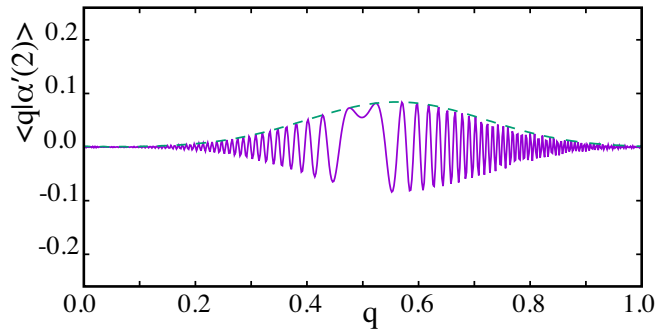}
\caption{{\bf Countering wave packet spreading}. The real (solid purple) and absolute (dashed green) values of $\braket{q|\alpha'(2)}$ are shown.  Applying the unitary transformation associated with the local linear canonical transformation of the inverse one-step stability matrix undoes wave packet spreading without shifting the position or momentum centroids and generates the second panel from the top of Fig.~\ref{fig3}.}
\label{fig4}
\end{figure}

The kicked rotor has a long history of providing a simple, powerful paradigm for both classical and quantum dynamical systems~\cite{Chirikov79, Izrailev90, Lakshminarayan97, Tomsovic07}.  Its quantum version has been experimentally realized with cold atoms in a kicked optical lattice for a variety of purposes~\cite{Moore95, Ammann98, Ryu06, Sajjad22}. 
The classical Hamiltonian is given by
\begin{equation}
\label{krg}
H(q,p) = \frac{p^2}{2} - \frac{K}{4\pi^2}\cos (2\pi q) \sum_{n=-\infty}^\infty \delta(t-n) \, .
\end{equation}
By choosing the kicking strength greater than roughly $2\pi$, it generates a strongly chaotic classical dynamics.  The resulting mapping equations for the version on the unit-periodicity phase space torus are
\begin{equation} 
\label{eq:two}
\begin{split}
& p_{n+1} =p_{n}-\frac{K}{2\pi }\sin (2\pi q_{n}) \  \pmod 1 \, , \\
& q_{n+1} =q_{n}+p_{n+1}  \qquad\qquad \pmod 1  \, .
\end{split}
\end{equation} 
The single step stability matrix,
\begin{equation}
\left( \begin{array}{c} 
\delta p_{n+1} \\  
\delta q_{n+1}
\end{array}\right) = {\bf M}_n 
\left( \begin{array}{c} 
\delta p_{n}  \\ 
\delta q_{n} 
\end{array} \right) \; ,
\label{deltas2}
\end{equation}
 is critical for the targeting $U_{M^{-1}}$ and given by
\begin{equation}
{\bf M}_n = \left( \begin{array}{cc} 
m_{11} & m_{21} \\ 
m_{12} & m_{22} \end{array} \right) = 
\left( \begin{array}{cc} 1 & -K \cos \left(2 \pi q_n \right) \\
 1 & 1-K \cos \left(2 \pi q_n \right) \end{array} \right) 
\label{deltas1}
\end{equation}
and with inverse given by
\begin{equation}
{\bf M}_n^{-1} = \left( \begin{array}{cc} 
m_{22} & -m_{21} \\ 
-m_{12} & m_{11} \end{array} \right) \ .
\label{deltas-1}
\end{equation}
The quantum dynamics are generated by a unitary or Floquet operator,
\begin{equation}
\widehat{U} = \exp\left( \displaystyle\frac{-i \widehat{p}^2}{2\hbar} \right) \; \exp\left[ \displaystyle\frac{iK}{4\pi\hbar^2}  \cos 2\pi\widehat{q} \right] \;.
\label{eq.4a3}
\end{equation}
In a configuration representation, $U_{jk} = \langle q_j \mid \widehat{U} \mid q_k\rangle$, with null Bloch phases it becomes
\begin{equation}
\resizebox{.9\hsize}{!}{$U_{jk} =\! \displaystyle\frac{1}{\displaystyle\sqrt{iN}} \exp\!\left[ \displaystyle\frac{i\pi (j\!-\!k)^2}{N} \right] \exp\!\left[ \displaystyle\frac{iNK}{2\pi}\! \cos\displaystyle\frac{2\pi k}{N} \right]$ ,}
\label{eq.4a4}
\end{equation}
where $N$ is the Hilbert space dimension, $j,k = 1,...N$, and Planck's constant is $2\pi\hbar = 1/N$.  The semiclassical limit of $\hbar \rightarrow 0$ is equivalent to $N\!\rightarrow \!\infty$.  The kicking strength $K\!=\!4\pi^2/5$ and Hilbert space dimensionality $N=500$ are chosen for Fig.~\ref{fig2}-\ref{fig4}.

The initial state $\ket{\alpha}$ reads
\begin{equation}
\label{gauss}
\braket{q_j | \alpha} = A(\hbar)\exp\!\left[\!- \frac{\left(q_j-q_\alpha\right)^2}{2\hbar} \!+\!\frac{i}{\hbar}p_\alpha \left(q_j\!-\!q_\alpha\right) \right]
\end{equation}
where $A(\hbar)$ is a normalization constant, and likewise for the target state
$\ket{\beta}$. This form shares momentum and position minimum uncertainty equally on the unit torus.  Thus, the wave packet's phase space analogy has a circular symmetry for all $\hbar$.  The area inside the wave packet's two standard deviation contour is equal to $h\ (=1/N)$.  The integer value $m$ of the position $q_j=m+j/N$ is chosen such that $\left| q_j-q_\alpha\right| \le 1/2$ for all $j$.  In this way, $p_\alpha = \braket{ \alpha| \widehat p |  \alpha}$  and likewise,  $q_\alpha = \braket{ \alpha| \widehat q |  \alpha}$. The selected wave packet centroids are $(q_\alpha, p_\alpha)=(0.5,0)$ and $(q_\beta, p_\beta)=(0,0)$, respectively.

\begin{table}[ht]
\centering
\caption{{\bf Targeting heteroclinic trajectory.} Values $n=1,...4$ correspond to position and momentum centroids of the propagated states in the four panels of Fig.~\ref{fig3}.}
\begin{tabular}{| c | c | c |}
\hline
 n & $q_n$ & $p_n$ \\
\hline
\hline
\ 0 \  & \ 0.50063152866 \  & \  0.00056701866 \ \\
\hline
1 & 0.50618488472 &  0.00555335605 \\
\hline
2 & 0.56055984292 &  0.05437495820 \\
\hline
3 & 1.08164081735 &  0.52108097443 \\
\hline
4 & 0.98601176250 & -0.09562905485 \\
\hline
\end{tabular}
\label{table1}
\end{table}

In such a strongly chaotic system the initial state spreads rapidly on a logarithmically short time scale, the Ehrenfest time, $\tau_E\!=\! \ln(S/\hbar)/\mu$; see Fig.~\ref{fig2}.  The classical action $S$ can be taken as unity for the unit torus kicked rotor, and the Lyapunov exponent, $\mu\!=\!\ln (K/2)$.  Here, $\tau_E\approx 4$, which implies ergodic statistical behavior by this time.  Further propagation appears random.

\begin{figure}[ht]
\begin{center}
\includegraphics[width=\columnwidth]{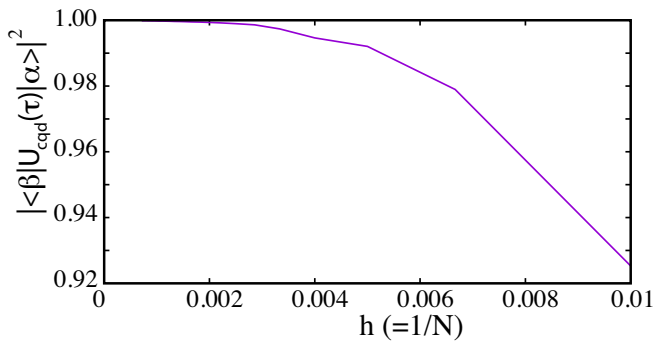}
\end{center}
\caption{{\bf Fidelity of quantum targeting for the  kicked rotor}.  The squared overlap of the target state with the controlled propagated state, i.e.~$|\braket{\beta|U_{cqd}(\tau=4)|\alpha(0)}|^2$, see Eq.~(\ref{unt}), is plotted  versus Planck's constant.  As $\hbar\rightarrow 0$, the error tends to vanish exponentially.  Dimensionalities ranging from $N=100$ to $1450$ are included.
}
\label{fig5}
\end{figure}

To illustrate quantum targeting for the kicked rotor, a heteroclinic trajectory segment of four iterations is chosen and given in Table~\ref{table1}.  The initial point is quite close to $(q_\alpha, p_\alpha)$ and the final point not quite as close to $(q_\beta, p_\beta)$.  Extending this trajectory segment to six iterations, one prior and one latter, would make using the before and after shifts, $U_s$, largely unnecessary in Eq.~\eqref{unt}; they are kept in this example.  With the wave packet spreading countered at each time step, as in Eq.~\eqref{unt}, it can be seen that the initial state follows the chosen heteroclinic trajectory exceedingly well; see Fig.~\ref{fig3}.  In fact, the absolute squared overlap of the evolving state following Eq.~\eqref{unt}, with Gaussian wave packets created using Eq.~\eqref{gauss} and the successive heteroclinic phase points, is equivalent to the accuracy shown in Fig.~\ref{fig5} ahead along the entire trajectory segment.

The important role of $U_{M^{-1}}$ is illustrated for the second iteration of the Floquet operator in Fig.~\ref{fig4}.  With $\ket{\alpha'(2)}$ the propagated ket before applying $U_{M^{-1}}$ and $\ket{\alpha(2)}$ afterwards, the effect of the inverse stability matrix of the single step of the heteroclinic trajectory is markedly relocalizing, in fact bringing the quantum state back to minimum uncertainty.  This technique is quite flexible in the sense that one could undo the stretching continuously in time or even double the time shown here by preemptively undoing the stretching of the coming next iteration or other choices.

As a function of a shrinking $2\pi\hbar=1/N$, Fig.~\ref{fig5} demonstrates the exponentially rapid decrease in the difference from unity of the overlap between the target state and the state propagated via quantum control, $U_{\rm cqd}(\tau)$, for the quantum kicked rotor.  This serves as a proof of principle and demonstrates the efficiency of our method.

There are two main sources of inaccuracies arising in this illustration, one due to the nonlinearities in the local dynamics and the other due to working with the periodic nature of the torus in this specific example.  With regards to the nonlinearities, it gives rise to curvature of the stable and unstable manifolds.  With regards to the torus, the Gaussian construction above, Eq.~\eqref{gauss},  is not explicitly periodic.  Both sources lead to vanishing errors as $\hbar\rightarrow 0$.

In conclusion, a generalization of targeting control in classically chaotic systems is constructed for quantum systems that is fully coherent (unitary).  Using the system's extreme sensitivity to initial conditions, counteracting the quantum spreading (scrambling) is demonstrated, and by means of a weak perturbation, the quantum system is put on an optimum track in order to arrive in an exponentially fast way and with high precision at a predetermined target state and given time. Hence, in a sense there are no free optimization parameters as everything is determined by the heteroclinic trajectory segment(s) chosen and its short time stability matrices.  The method is flexible in the sense that any trajectory segment of the unperturbed chaotic classical system can be selected and followed, plus the unwinding of the wave packet uncertainty spreading can be accomplished in multiple ways. In contrast to classical targeting control, the most critical element in the quantum case is the local unitary transformation $U_{M^{-1}}$. 

The underlying protocol, comprising time sequences of tailored $U_{M^{-1}}$, can be regarded as a sophisticated Floquet engineering method.  Furthermore, if the initial state is specifically chosen as the target state, stabilized periodic quantum dynamics (with any desired period) can be achieved.

Our proposed method is, by construction, rather general and may serve as a conceptual platform for a broader range of applications in various branches of quantum control.  In particular, the ideas presented here naturally extend to coherent states in bosonic many-body systems as linear canonical transformations are related to squeezing~\cite{Han88, Walls83}, and that is left for future work.

\acknowledgments
We thank Mathias Steinhuber for scientific discussions and help with devising Fig.~\ref{fig1}.  We acknowledge support by the Deutsche Forschungsgemeinschaft (DFG, German Research Foundation), project Ri681/15-1, within the Reinhart-Koselleck Programme and Vielberth Foundation for financial support.  

\bibliography{classicalchaos,furtherones,general_ref,molecular,quantumchaos,rmtmodify,manybody}

\end{document}